\documentstyle[12pt]{article}
\begin{document}
\def\be{\begin{equation}}
\def\ee{\end{equation}}
\def\bc{\begin{center}}
\def\ec{\end{center}}
\def\beq{\begin{eqnarray}}
\def\eeq{\end{eqnarray}}
\def\beqd{\begin{eqnarray*}}
\def\eeqd{\end{eqnarray*}}
\def\nin{\noindent}
\def\lra{$\leftrightarrow$ }

\def\etaten{$\eta_{10}$}
\def\he{$^4$He}
\def\dhe{D$+^3$He/H}
\def\rhoe{$\rho_e$}
\def\Pe{$P_e$}
\def\drhoedt{$d\rho_e/dT_{\gamma}$}
\def\tnu{T_{\nu}}
\def\rhonu{\rho_{\nu}}
\def\nnu{N_\nu}
\def\dnnu{\Delta \nnu}
\def\tg{T_{\gamma}}

\rightline{{\bf CWRU-P10-95}}
\rightline{{\bf astro-ph/9509126}}
\rightline{September 1995}
\baselineskip=16pt
\vskip 0.5in
\begin{center}
\bf\large {BIG BANG NUCLEOSYNTHESIS CONSTRAINTS ON
PRIMORDIAL MAGNETIC FIELDS
}
\end{center}
\vskip0.2in
\begin{center}
Peter J. Kernan, Glenn D. Starkman
and
Tanmay Vachaspati
\vskip .1in
 {\small\it Department of Physics\\
Case Western Reserve University\\ 10900 Euclid Ave., Cleveland, OH
44106-7079}
\vskip 0.4in
\end{center}
\centerline{{\bf Abstract}}
\noindent
We reanalyze the effect of magnetic fields in BBN, incorporating
several features which were omitted in previous analyses.
We find that the effects
of coherent magnetic fields on the weak interaction rates and the
electron thermodynamic functions (\rhoe, \Pe, and \drhoedt ) are
unimportant in comparison to the contribution of the magnetic
field energy density in BBN.
In consequence the effect of including magnetic fields in
BBN is well approximated numerically by treating the additional
energy density as effective neutrino number. A conservative
upper bound on the primordial magnetic field,
parameterized as $\zeta=2eB_{rms}/(T_\nu^2)$, is $\zeta \le 2$
($\rho_B < 0.27 \rho_\nu$). This bound can be stronger than
the conventional bound coming from the Faraday rotation measures
of distant quasars if the cosmological magnetic field is
generated by a causal mechanism.
\newpage

\section{Introduction}

It has frequently been suggested that a significant cosmological
magnetic field is present \cite{parker, zeldovich, kroneborg}
and over the last few years several proposals have been made
how to generate such a field \cite{proposals}.
Big bang nucleosynthesis (BBN) provides us with
a tool that can be used to probe physics in the early universe and
as such is invaluable in the study of primordial magnetic fields.
In fact, BBN appears to be the only tool presently available that can
explore magnetic fields in the pre-recombination universe
and hence the physics
that goes into generating the magnetic field.

The effect of magnetic fields on BBN was discussed in the
early pioneering works of
Greenstein \cite{greenstein} and Matese and O'Connell \cite{matese}.
The conclusions and understanding in these papers
were broadly correct. But now, with advances in our understanding
of BBN and faster computers, we can make tighter quantitative estimates
of the light
element abundances (\he , D and $^3$He). The issue has also been
examined in the recent literature,
first by Cheng, Schramm and Truran (CST) \cite{CST} and subsequently
by Grasso and Rubinstein (GR) \cite{GR}. Our analysis differs from
these calculations in that we have included several factors omitted
in these works, corrected some apparent errors and resolved some of
the conflicts. We also provide analytic arguments based on the
Euler-Maclaurin expansion to deduce the changes in various quantities
due to the magnetic field.

There are two primary effects of magnetic fields on BBN :
(i) the magnetic field energy density contributes to
the cosmological expansion rate, and, (ii) the electron
phase space changes due to the magnetic field. {\it A priori},
we cannot say if the dominant effect that determines any changes to the light
element abundances is (i) or (ii), so it is necessary to
go through the full analysis and to evaluate the magnitudes of both
effects. The determination of the effects of the electron phase
space change are quite involved and it is here that we disagree
with CST and GR. However, after the full evaluation is carried out,
the dominant effect of the magnetic field turns out to be its direct
contribution to the cosmological expansion rate.
The computationally expensive contributions to the weak rates and
the electron thermodynamic functions can be neglected, and in practice
the numerical BBN computation can be avoided altogether by expressing
the bound on the magnetic field energy
density as an equivalent bound on the effective
number of neutrinos, a number commonly derived in the BBN literature.
This should be contrasted with the claims of GR who find that the
change in \rhoe\ dominates, and CST who find that the changes in
the weak interaction rates dominate. We agree with the original
conclusions of Matese and O'Connell.

Although a rigorous comparison of the effects (i) and (ii) can only be made
after a systematic evaluation, one way to understand why
the effect on the expansion rate is more important than the change
in the interaction rates is to note that the expansion rate changes
in proportion to $B^2$ while the interaction rates change in proportion
to $\alpha B^2$, with $\alpha \simeq 1/137$ being the fine structure constant.
Any other numerical factors occuring in the interaction rates turn out not
to be large enough to overcome the suppression by $\alpha$.

Our analysis assumes a magnetic field that scales as:
\be
B \propto R^{-2} \equiv b\tnu^2
\label{bdef}
\ee
where $R$ is the scale factor of the universe and $\tnu$ the neutrino
temperature. Our analysis
also assumes that the cosmological expansion is well described
by the Friedmann-Robertson-Walker (FRW) model. The latter assumption
is only justified if the magnetic field is sufficiently tangled
on scales smaller than the horizon - otherwise we would need
to consider the anisotropic expansion of the universe.
At the same time, the assumption in  (\ref{bdef})
is valid only on scales large enough for the plasma conductivity
to keep the magnetic field frozen-in. (Today this frozen-in scale is
of order $10^{13}$ cms.) Recent work by Jedamzik et. al.
\cite{scale} considers the magnetohydrodynamical evolution of the
system and concludes that the magnetic field can dissipate on yet
larger scales which would correspond to scales of several megaparsecs
today. The BBN constraints that we derive on
primordial magnetic fields are therefore only valid for magnetic fields
that are coherent on scales larger than the scale on which magnetic
fields can dissipate and smaller than the horizon scale. In particular,
magnetic fields generated during inflation are likely to be coherent
on super-horizon scales and have not been fully dealt with by our
analysis.

As the dominant effect of magnetic fields on BBN is due to the
contribution of the field to the energy density, it will be convenient
to express the energy density in magnetic fields
in terms of the number of equivalent additional neutrino species.
The energy density contributed by the magnetic field is
\be
\rho_B = {B^2\over 8\pi }.
\label{rhoB}
\ee
where, $B^2$ is the volume average of $|{\vec B}(\vec x )|^2$.
The contribution of $N_\nu$ light ($m \ll 1 MeV$) neutrino
species to the energy density of the universe is
\be
\rhonu = {{7\pi^2} \over {120}} N_\nu  \tnu^4.
\label{rhonu}
\ee
and so $\rho_B$ may be written
as a number of equivalent additional neutrino species
\be
\Delta N_{\nu}^B = {15 \over 7\pi^3}b^2,
\ee
where $b = B/T_\nu^2$ is  the constant which we  seek to constrain.
This gives
\be
\rhonu+ \rho_B = (N_\nu + \Delta N_\nu^B) \rhonu (N_\nu=1).
\ee
For standard BBN $N_\nu=3$.
Krauss and Kernan (KK) argue that a
very conservative bound on $\dnnu$ is $\dnnu < .8$ \cite{KK3},
leading to the bound $b < 3.4$.

The bound in KK on
$N_{\nu}$ is derived by requiring that at least 5\% of 1000 Monte
Carlo runs  simultaneously satisfy the abundance constraints
\dhe $< 10^{-4}$ and $Y_p < .25$, for
any value of $\eta$, the baryon to photon ratio, for a given
number of neutrinos. The random variables in the
the BBN Monte Carlo uses 11 nuclear reaction rates
and the neutron lifetime as random variables.
The result, $N_{\nu} < 3.8,$ is conservative in that the
abundance limits assumed
allow for generous systematic errors in the observed $^4$He/H
abundance in HII regions,
and the likelihood of anomalous presolar abundances of
D and $^3$He.

In the following sections we
lay the physics groundwork for our bound on primordial magnetic
fields.
First we consider the effects
of the {\em B} field on the electron phase space, then the
resultant changes in the weak interaction rates and the
time-temperature relation. Finally we tie these all together
and numerically solve for the neutron fraction as a function
of temperature.

\section{Electron Phase Space}

The dispersion relation for the electron in the presence of
a uniform magnetic field is
\be
E= (p_e^2 + m_e^2 + eB(2n+s+1))^{1/2}
\label{disp}
\ee
\nin with {\em e} the electron charge,
{\em B} the magnetic field strength,
and {\em n} the principal, and {\em s} the magnetic
quantum number of the Landau level.
This relation is valid for a magnetic field that is small compared
to the critical field $B_c = m_e ^2/e$ and has only been derived
at zero temperature. For stronger fields and at high temperatures
we expect further corrections to this formula. For example, the
next order correction would be to include the anomalous magnetic
moment of the electron. However, the analyses available in the
literature \cite{GRHRW} indicate that these higher order corrections
generally have a very small effect at the
magnetic field strengths and temperatures of interest and so we will
only consider the dispersion relation in (\ref{disp}) throughout
this paper. The full treatment of the problem
is computationally impractical until an
analytic approximation is developed for $E(n,s,p,B)$.

After summing over the electron spin in (\ref{disp})
the zero-chemical potential phase space of electrons appearing in
weak interaction rates and electron thermodynamic functions changes
according to
\be
2{d^3p\over (2\pi)^3 } f_{FD}(E_0)\ \ \longrightarrow \ \
\sum_{n=0}^{\infty} (2-\delta_{n0}) {dp_e \over 2\pi}
{eB\over 2\pi} f_{FD}(E_B,T)
\ee

\nin with $E_B^2=p_e^2+m_e^2+2eBn$ and
$f_{FD}(E,T)$ the Fermi-Dirac distribution at temperature T:

\nin
\be
f_{FD}(E,T) = {1\over 1 + exp(E/T)}.
\ee
We have neglected the chemical potential, $\phi$ of the electrons
since $\phi/\tg$ is of order the baryon to photon
ratio, $\eta$, which is small( {\cal O}($10^{-10}$));
hence any $B$ and $\phi$-dependent
correction to the standard result would be second order in small quantities.
The electron thermodynamic functions
are given by the usual prescription of integrating the relevant quantity
over phase space.
Thus

\beq
n_e &=& {eB\over (2\pi)^2}\sum_{-\infty}^{+\infty} (2-\delta_{n0})
\int_{-\infty}^{+\infty}
f_{FD}(E_B,\tg) dp_z \label{nume}\\
\rho_e &=& {eB\over (2\pi)^2}\sum_{-\infty}^{+\infty} (2-\delta_{n0})
\int_{-\infty}^{+\infty}
E_B f_{FD}(E_B,\tg) dp_z \label{rhoe}\\
P_e &=& {eB\over (2\pi)^2}\sum_{-\infty}^{+\infty} (2-\delta_{n0})
\int_{-\infty}^{+\infty}
{E_B^2 - m_e^2 \over 3 E_B} f_{FD}(E_B,\tg) dp_z \label{Pe}\\
\eeq

\nin $n_e$ does not appear in the time-temperature relation, but
is useful for illustrative purposes, e.g. the perhaps
counter-intuitive result that the electron number density
increases as the
magnetic field increases.
This is easy to show by expanding (\ref{nume}) in an Euler-MacLaurin
expansion.
\beq
n_e &\simeq& {\tg^3 \over 4\pi^2} \bigl[
  4 \int_0^\infty {\rho^2 d\rho \over
	1 + e^{\sqrt{\rho^2 + m_e^2/\tg^2}}}  \nonumber \\
  &+&  {(\zeta r_{\nu\gamma})^2 \over 24}
 	\int_0^\infty {d\eta\over \sqrt{\eta^2 + m_e^2/\tg^2}}
{1\over 1 + \cosh\sqrt{\eta^2 + m_e^2/\tg^2}} \label{neEulerMcL}\\
&+& {\cal O}((\zeta r_{\nu\gamma})^4) \bigr] \nonumber
\eeq
where $\zeta = 2 e B/T_\nu^2 = 2 e b$, and
$r_{\nu\gamma} = (T_\nu/\tg)^2$.
Although we will quote limits on $b$ or $\zeta$, which are constants,
we will also use $z=\zeta r_{\nu\gamma}$,  which is what usually appears
in the equations.
Clearly from (\ref{neEulerMcL}),
\be
\lim_{B\to 0}{dn_e \over dB} \ >  \ 0
\ee
The interpretation is that there are two effects of increasing the
magnetic field strength on the electron phase space -- 1) the energy
cost of populating $n > 0$ levels becomes greater due to the
larger spacing of the Landau levels, and,
2) the areal element in the phase
space decreases due to the decreased cross-sectional area of
each Landau level. The second effect is stronger than the first
and causes the number density of electrons to increase with $B$.
For $\rho_e$, the sign of the effect is less clearcut (indeed it depends on
the temperature):
\beq
\rho_e &\simeq& {\tg^3 \over 4\pi^2} \bigl[
  4 \int_0^\infty {\rho^2 \sqrt{\rho^2 + m_e^2/\tg^2} d\rho \over
	1 + e^{\sqrt{\rho^2 + m_e^2/\tg^2}}} \nonumber\\
  &+& {z^2 \over 12}  \int_0^\infty d\eta
\left( {1\over 2 (1 + \cosh\sqrt{\eta^2 + m_e^2/\tg^2})} -
{1/\sqrt{\eta^2 + m_e^2/\tg^2} \over
	1 + e^{\sqrt{\rho^2 + m_e^2/\tg^2}}} \right)\label{rhoeEulerMcL}\\
&+& {\cal O}(z^4) \bigl];\nonumber
\eeq
The first term inside the square brackets gives the electron energy density
when $B=0$. The second term is the correction to the energy density due
to the presence of a magnetic field and can be negative or positive,
depending on the temperature $\tg$.

Assuming that the integral in the $z^2$ term of (\ref{rhoeEulerMcL})
is  ${\cal{O}}(1)$,
it is clear  that the change in
the energy density  of electrons in the field
is smaller than the pure magnetic field contribution  to the energy density.
Since the former is due to coupling of the magnetic field to the electrons,
it is proportional to $z^2 \propto \alpha B^2$ times
small numerical factors of order $0.01-0.1$.
This is smaller by $\cal{O}(\alpha)$ than the $B^2/8\pi$  term.
We expect the phase space corrections to the
weak interaction rates to also be of order  $z^2$
times small numerical factors.

In \cite{GR}, only the changes in \rhoe\ were taken into account;
but the changes in \Pe\ and \drhoedt\ must also be taken into account
since they appear in the BBN time-temperature relation.
We will spare the reader the explicit formula for \drhoedt\ which
is quite a lengthy expression with many extra terms involving $dB/d\tg$.

\section{Neutron Fraction}

The binding energy per nucleon of helium-4, the most strongly bound
light nucleus, is  7.1 MeV.
Yet, well below this temperature the abundance of $^4$He compared to
hydrogen is quite small,
because the high energy tail of the photon
distribution breaks up any deuteron ($Q = 2.2$MeV) which forms.
At a (slightly $\Omega_B h_0^2$ dependent ) temperature of $\tg \simeq .07 MeV$
the high energy tail of the photon distribution
is no longer energetic enough to break up all the deuterons.
When this ``deuterium bottleneck" is passed,
all the available neutrons are swiftly incorporated into \he,
with trace amounts of other light nuclei, D, $^3$He and $^7$Li produced.
To a very good approximation, $Y_p$, the primordial \he\ mass fraction is
given by twice the neutron fraction at the breaking of the
deuterium bottleneck.
Thus to discover the importance of the effects
of the corrections to the BBN dynamics,
for a first pass it is sufficient to examine the evolution of the
neutron fraction with all strong interations turned off.

With respect to comparison with observed abundances
the $^4$He production is the most likely
to be noticeably affected by the changes in BBN dynamics
in the presence of the magnetic field.
This is because the \he~ observations are sensitive at the $5\%$ level or
better,
whereas all the other observations are sensitive only at the 50-100\% level.
Also, changes induced in the nuclear reaction rates
when they are important ($\tg \leq 0.07$MeV)
are likely to be much smaller than
those induced in the weak interaction rates
when they are important ($\tg \gg 0.07$MeV),
both because the electron magnetic moment
is much larger than the nuclear magnetic moment,
and because the magnetic field decreases as $T_\nu^2 \simeq \tg^2$,
and is therefore small by the time the nuclear reactions take place.
Finally, the other elements are sensitive
to changes in the time-temperature relation due to the magnetic field
only because of the very small change
in the temperature at which their number density falls out of
quasi-equilibrium,
whereas the neutron fraction at the breaking of the deuterium bottleneck
(and hence the \he\ abundance)
is sensitive to the change in the equilibrium neutron fraction during
weak equilibrium,
the change in the departure of the actual neutron fraction
from it equilibrium value,
the change in the freezeout temperature of the weak interactions,
the change in the fraction of neutrons which decay
between weak freezeout and the breaking of the bottleneck,
and the change in the temperature at which the deuterium bottleneck breaks.

The neutron fraction evolves according to
\be
{dX_n \over dt} = - \lambda_{np} X_n + \lambda_{pn}(1 - X_n)
\label{dxndt}
\ee
where $\lambda_{np}$ is the rate
per nucleon of neutron conversions to protons
and $\lambda_{pn}$ is the rate of proton conversions to neutrons.
\be
\lambda_{np} = \lambda_{n\to pe^-\nu} + \lambda_{n e^+\to p\nu}
+ \lambda_{n\bar{\nu}\to pe^-}
\ee
with a corresponding formula for $\lambda_{pn}$.
For the numerical calculation we left out the 3-body reaction,
$pe^-\nu\to n$ which never becomes important.
If the weak interactions were in equilibrium, which would be the case
if the interaction rates were much faster than the Hubble expansion rate,
then the neutron fraction would assume its equilibrium value
\be
X_n^{eq} = {\lambda_{pn} \over\lambda_{np} +  \lambda_{pn} }
\label{Xneq}
\ee

The weak interaction rates at these low energies are given simply
by integrals of $p_e E_e E_\nu^2$ over the available phase space.
For example,
\be
\lambda_{n\to pe^-\nu} = {G_F^2 \tg^2 (g_V^2 + 3g_A^2) \over (2\pi)^3}
z \sum_{n=0}^{n_c} (2 - \delta_{n0})
\int_{0}^{p_m} dp_z E_{\nu}^2 g(E_e/\tg)g(E_{\nu}/\tnu)
\label {lambda}
\ee
\nin Here $G_F$ is the Fermi decay constant,
$g_V$ and $g_A$ are the vector and axial-vector coupling constants.
The electron energy is given by
\be
E_e = (p_e^2 + m_e^2 + 2eBn)^{1/2}\ ,
\ee
the neutrino energy by
\be
E_{\nu} = \Delta - E_e,
\ee
with $\Delta = 1.293 MeV$ being the neutron-proton mass difference.
The sum in  (\ref{lambda})
is taken to a maximum of $n_c = $[$(\Delta^2-m_e^2 )/ 2eB$]
(where [$x$] is the largest integer $\leq x$),
and the momentum integral is up to
$p_m = \sqrt{\Delta^2-m_e^2-2neB}$.
$g$ is the Fermi blocking factor and is given by:
\be
g(E/T) \equiv 1 - f_{FD}(E/T) = (\exp(-E/T) + 1)^{-1}.
\ee
The other weak interaction rates have a similar form,
with appropriate factors of $f_{FD}$ and $g$ in the integrands,
and with the limits on the sum and integral given by the requirement
of positive energy for the electron and neutrino.
For example,
\be
\lambda_{n\nu\to pe^-} = {G_F^2 \tg^2 (g_V^2 + 3g_A^2) \over (2\pi)^3}
z  \sum_{n=0}^{\infty} (2-\delta_{n0})
\int_{0}^{\infty} dp_z E_{\nu}^2 g(E_e/\tg)f(E_{\nu}/\tnu)
\ee

With the exception of beta decay we find that the magnetic
field effects on the electron phase space decreases all the
weak rates. For small $z$ we can see this once again using
the Euler-McLaurin expansion:
\beq
\lambda_{n\nu\to pe^-} &=& {G_F^2 \tg^2 (g_V^2 + 3g_A^2) \over (2\pi)^3}
\bigl( 2\int_0^\infty dx \int_{p_{min}(x)}^{\infty} G_-(E_e(p_z,x)) dp_z
\label{lambdannu}\\
&-& {z^2 \tg^2 \over 12}  \int_{p_{min}(0)}^{\infty}
\left[ {1\over E_e}{d G_-(E_e)\over dE_e}\right]_{x=0} dp_z
\bigr) \nonumber
\eeq
where
\be
E_e(p_z,x) = (p_z^2 + m_e^2 + x)^{1/2},
\nonumber
\ee
\be
p_{min}(x) = (\Delta^2 - m_e^2 - x \tg^2)^{1/2}
\nonumber
\ee
and
\be
G_-(E_e) = (E_e - \Delta)^2 {1\over 1+ e^{(E_e -\Delta)/T_\nu}}
			    {1\over 1+ e^{-E_e/\tg}}.
\ee
Thus
\beq
\left[ {1\over E_e}{d G_-(E_e)\over dE_e}\right] &=&
{E_e - \Delta\over E_e} {1\over 1+ e^{(E_e -\Delta)/T_\nu}}
		 {1\over 1+ e^{-E_e/\tg}}\\
&\times&\left[2 - {E_e - \Delta\over T_\nu}
         {1\over 1+ e^{-(E_e -\Delta)/T_\nu}}
	+ {E_e\over \tg} {1\over 1+ e^{E_e/\tg}}\right] \ .
\nonumber
\eeq
We see that only the second term in the expression in square brackets
is negative and
that (for $x=0$), by the time $E_e$ is large enough for the term to
dominate over the other two terms in the square brackets,
the prefactor is exponentially suppressed.
The coefficient of the $z^2$ term of (\ref{lambdannu}) is therefore negative,
and the B-field decreases this rate.
Similar arguments can be applied to the other $2\to 2$ rates.
Though the conclusions are not always so clear analytically,
our numerical studies show
a decrease in all the 2-body rates for $B > 0$.

For beta decay the momentum cutoff introduces
an $n$ cutoff, and with $B>B_c= (\Delta^2 - m_e^2) / 2e$
only the $n=0$ term survives in the sum.
Thus it is easily seen that $\lambda_{n\to pe^-\nu}$
increases linearly with $B$ for $B\ge B_c$.
We find numerically that it also increases (though not linearly)
for $B < B_c$.

There are three competing
effects for the weak interactions.
At high temperatures the weak rates are slower.
This by itself does not imply anything about the equilibrium
neutron fraction,  since the neutron fraction
(see  (\ref{Xneq})) depends on a ratio of rates;
but,
because  $\lambda_{pn}$ decreases more than $\lambda_{np}$,
there is a small decrease in the equilibrium neutron fraction
(see Figure 1). This is the first effect.
However, below about 1 MeV, the weak interaction rates are  too slow
to maintain this equilibrium.  Since $X_n^{eq}$ is decreasing with
temperature, the actual neutron fraction lags behind, and is
therefore always greater than, the equilibrium value.
Because the B field lowers the interactions rates,
$X_n$ is less able to  track $X_n^{eq}$, and so
lags behind the dropping equilibrium neutron fraction more than usual.
This second effect tends to increase the neutron fraction at a
given temperature.
Finally at later times,
$\tg$ \raisebox{-3pt}{\mbox \small {$\stackrel{<}{\sim}$}}$.2 MeV$,
neutron decay dominates the weak interaction rates.
The $B$ field causes the neutron decay rate to increase,
lowering the neutron fraction.
However, since neutron decay is the dominant reaction
only at lower temperatures,
the $B$ field has already decreased significantly,
and the change in the decay rate is negligible.

If these changes in the reaction rates were the only effects,
then the decline in the equilibrium fraction and
the increase in the neutron decay rate at low temperatures
would outweigh the increased departure from equilibrium
(due to the  lower reaction rates)
at high temperatures.
The net result would  be a lower neutron fraction and hence a
lower \he~ abundance
as can be seen from the dotted line in Figure 2.

However, as we shall discuss in detail below, these are not the dominant
effects.

\section{Time-Temperature Relation}

To solve for BBN abundances  we must follow
not just the evolution of the neutron fraction,
but also the evolution of the photon and neutrino temperatures with time.
Initially, $\tg \geq 5 MeV$,
the neutrinos and electromagnetic fluids are assumed to be
in perfect thermal contact and $\tg^i=\tnu^i$.
Later, the photon and neutrino temperatures evolve differently
due to electron-positron annihilation \cite{SW,KT},
which heats the photons, but not the neutrinos.
This assumes zero net energy exchange between the neutrino and electromagnetic
fluids,
a very good approximation for BBN \cite{DT}.
Assuming that the neutrinos are decoupled well before this annihilation takes
place,
$T_\nu$ evolves as the inverse scale factor

\be
R\tnu = constant.
\label{rtnu}
\ee
For the photons we use energy conservation in the electromagnetic plasma.

\be
d[(\rho_{em} + P_{em})R^3] = R^3d P_{em}, \nonumber
\ee
\nin with $\rho_{em} = \rho_e + \rho_{\gamma}$ and
$P_{em} = P_{\gamma} + P_e$.
It follows that
\be
{d\rho_{em}/ d\tg} = -3 (\rho_{em} + P_{em}) R^{-1}{dR/  d\tg} \label{econs}
\ee
\nin The magnetic field energy density is not to be included in (\ref{econs}).
This would be appropriate only if the large scale {\em B} field was
in thermal equilibrium with the electromagnetic fluid,
and hence evolved according to $\tg$ rather than the inverse scale factor.

The Hubble expansion law for the radiation dominated
early universe is
\be
H \equiv {1 \over R}{dR\over dt} = \sqrt{8\pi G \rho_T \over 3}
\label{hubble}
\ee

\nin with $\rho_T$ representing all forms of energy density in the
universe. (Note that the magnetic field energy is included here
but not in (\ref{econs})).  From  (\ref{econs}) and (\ref{hubble})
we obtain the time-temperature
relation which we are seeking,
\be
{d\tg\over dt}= -3 H {\rho_{em} + P_{em} \over d\rho_{em}/d\tg}.
\label{ttemp}
\ee
\nin The analogous equation for the neutrino temperature, from
(\ref{rtnu}) and (\ref{hubble}) is

\be
{d\tnu \over dt} = -H\tnu.
\label{ttnu}
\ee
The complication in solving (\ref{ttemp}) in the case of a magnetic field
is that $\rho_e$ depends both on $\tg$ and on $B = b T_\nu^2$,
$\rho_e = \rho_e(\tg,T_\nu;\zeta)$.
We therefore write
\beq
{d\rho_e \over d\tg} &=& {\partial \rho_e \over \partial \tg}
      +   {\partial \rho_e \over \partial T_\nu} {d T_\nu \over d\tg}
\label{2drho} \ .
\nonumber \\
\eeq
If we now define
\be
{\partial \rho_e \over \partial \tg}
\equiv h(\tg,\tnu)\ , \ \ \
{\partial \rho_e \over \partial T_\nu}
\equiv {1 \over \tnu} j(\tg,\tnu)
\ee
we can rewrite
(\ref{ttemp}) Using (\ref{ttnu}) and (\ref{2drho}) as,
\be
{d\tg\over dt}= -3 H {\rho_{em} + P_{em} \over d\rho_{\gamma}/d\tg
+ h} (1 - {j \over 3 (\rho_{em} + P_{em})}).
\label{2ttemp}
\ee

\nin The right-hand sides of equations (\ref{ttnu}) and (\ref{2ttemp})
depend only on $T_\nu$, $\tg$, and the constant $\zeta$
(or equivalently $b$) and can be integrated together to obtain
$T_\nu(t)$ and $\tg(t)$.

Although the details must be obtained numerically,
the basic picture is clear.
The increased energy density due to the $B^2 /8\pi$  term means
that at any given temperature ($\tg$ or $T_\nu$)
the universe is younger and expanding faster than in the standard scenario
($B=0$).
This has two important consequences.
First, the weak interaction rates, which are already
slower than in the $B=0$ case
have an even harder time maintaining the equilibrium neutron fraction.
Since the equilibrium neutron fraction is dropping,
this increased lag will cause the actual neutron fraction to be higher than for
$B=0$.
Second, the time between weak interaction freezeout and
the breaking of the deuterium bottleneck is shortened.
Thus there is less time for the extra neutrons to decay than if $B$ were zero.
This also acts to increase the final neutron fraction.
Although there is a lowering of the temperature at which
the bottleneck is broken, this is a second order effect \cite{ESD}
and the change in this temperature turns out to be
only about $0.1\%$ and is therefore ignored.

The net result is that
if one takes into account only the change  in the time-temperature relation,
the final neutron fraction rises
and so does the \he\  abundance.

\section{Results}

We are left to determine whether the neutron fraction
rises because of the change in the time-temperature relation,
or falls because of the changes in the weak interaction rates.
We find that the effects of the change in the time-temperature relation
from the magnetic field energy density
are much larger than those in the time-temperature relation or
the weak interaction rates due to the changes
in the electron phase space.
We have not been able to demonstrate this convincingly using only analytic
methods;
however, basically it is because the changes in the phase space
are proportional to $\alpha B^2$, whereas the change in the energy density
is proportional to simply to $B^2$ (with no $\alpha$).

We integrated the coupled differential equations
(\ref{ttnu}), (\ref{2ttemp}) and (\ref{dxndt})
to obtain the photon and neutrino temperatures and the neutron fraction
as a function of time.
We ran the computer code
from  $\tg= 5$MeV to $.07 MeV$, when the deuterium bottleneck is broken
and almost all the remaining neutrons are converted to \he.
We did this for several values of $\zeta = 2 e B/T_\nu^2$.

In Figure 2 we display the evolution of the neutron fractions for $\zeta=0$,
and for $\zeta=2$ in each of five different limits.
The long dashed curve shows the $\zeta=0$ case;
the solid curve shows $X_n$ with the full calculation;
in the dotted curve we have turned on only
the dependence of the weak interaction rates on the magnetic fields;
in the short-dash curve we have only turned on the dependence on the magnetic
fields
only of the electron thermodynamic functions;
and with the dash-dotted curve we have only turned on the
dependence of the expansion rate on the energy density of
the magnetic field.
It is clear that the latter is by far the dominant effect.
The change in $X_n(\tg=.07 MeV)$ from the canonical case, $B=0$, is
$-2.0 \times 10^{-5}$ for weak interactions only,
$<5.0\times 10^{-6}$ for electron thermodynamics only
and $+4.8\times 10^{-3}$
for magnetic field energy density only.
Thus the limit on B is equivalent to one on the number
of neutrinos, and for $N_{\nu} \le 3.8$, our limit on $B$ is
given by,
\be
\zeta \le 2
\label{result}
\ee
In units of the neutrino energy
density this can also be expressed as
\be
\rho_B \le .27 \rho_\nu ,
\ee
where $\rho_{\nu}$ is the energy density contributed
by the standard 3 light ($<< 1$ MeV) neutrinos.

An additional point of interest is that
the change in the electron thermodynamics during the era of
$e^+e^-$ annihilation causes a slight decrease in $\tnu/\tg$ with
respect to the canonical value, $(4/11)^{1/3}$.
This can be understood heuristically
as follows. The magnetic field increases the electron number density.
This is equivalent to some fraction of an
additional degree of freedom residing in the electromagnetic fluid
in the early universe. When the $e^+e^-$ pairs annihilate the additional
entropy is transferred to the photons, thus ``heating'' them slightly
more with respect to the neutrinos. For the values of the magnetic
field allowed by BBN the change is small enough that it has little
effect on other cosmological parameters which depend on the neutrino
temperature, such as the mass of a light neutrino required for
$\Omega_{\nu}=1$. If very large magnetic fields were allowed,
for example, with $\zeta = 8.0\ (16.0)$, we find
\beq
{\Delta (\tnu/\tg) \over \tnu / \tg } = -1.7\% \ (-5.6\%),
\label{result2}
\eeq
and this effect becomes significant.

In Figure 3 we display the evolution of the neutron fraction
as a function of $\tg$ from $\tg=3$ to $\tg = 0.07$ for $\zeta=.1,2,4$.
The points on the curve where $t = .1, 1, 10$ and $100$ seconds are marked.
The two dominant effects which we described are manifest.
At high temperature, the $2 \to 2$ weak interactions are dominant.
The higher the B field the faster the universe expands at a given $\tg$,
and the more the actual neutron fraction lags behind
the decreasing equilibrium abundance.
At low temperatures, neutron decay is dominant.
The higher the B field, the sooner the universe
cools sufficiently to break the deuterium bottleneck,
so the sooner neutron decay terminates and the remaining neutrons
are converted almost instantaneously to \he.  This is seen by the
earlier termination of the $X_n$ line.  Both these effects lead
to a higher neutron fraction.

\section{Discussion}

Our analysis of the effects of magnetic fields on BBN have
allowed us to place constraints on the field strength present
at the BBN epoch. The analysis is based on the following
assumptions:

{\em a}) The cosmological model is an FRW model. This means that
the magnetic field does not lead to anisotropic expansion
of the universe. For this the magnetic field should not be
coherent on the scale of the horizon such as happens in proposals
that generate magnetic fields from inflation. However, it seems
likely that if the magnetic fields are not too strong, the
constraint we have derived using an isotropic FRW universe
should still apply. If there are significant departures from
FRW universes, the magnetic field would be constrained by
measurements of the isotropy of the universe stemming from
measurements of the microwave background\cite{madsen}.

{\em b}) The magnetic field is assumed to have a scaling given
by  (\ref{bdef}) and hence we have assumed that it does
not dissipate.
The shortest scale that does not dissipate has
been argued to be given by the magnetic Jeans length \cite{scale}.
Another scale that is probably relevant is the scale on which
the magnetic fields are frozen in due to the high electrical
conductivity of the cosmological plasma \cite{parker, zeldovich}.

{\em c}) The dispersion relation is assumed to be given by  (\ref{disp})
which is known to be incorrect when the magnetic field is strong
(compared to $m_e ^2 /e$)
or at high temperatures. But current analyses \cite{GRHRW}
indicate that the
corrections are small in the range of temperatures and magnetic
field strengths that are relevant for us.
We assume that these corrections will not change our result
significantly. This assumption can be removed once we have a better
{\it analytic} grasp of the corrections to the dispersion relation as
it would be computationally infeasible to use the high magnetic field
strength and temperature dispersion relation that is known in its
present form.

Within the framework of these assumptions, we have derived a BBN
constraint on the present cosmological
volume averaged magnetic field strength $B_{rms}$ (equivalently, on the
magnetic field energy density) defined by
$$
B_{rms}^2 = {1 \over V} \int_V d^3 x |{\vec B} (\vec x ) |^2 \ .
$$
Using  (\ref{result}) and the evolution equation (\ref{bdef}),
the constraint is:
\beq
B_{rms} <  10^{-6} G \ .
\label{constraint}
\eeq
We would like to emphasize that this constraint is a local constraint
and not a constraint on the magnetic field strength at any particular
coherence scale.

An oft quoted constraint on primordial magnetic fields comes from
the Faraday rotation measures of distant quasars. This constraint
is
\beq
\bar B < 10^{-9} G
\label{faraday}
\eeq
where $\bar B$ is the line-averaged magnetic
field:
$$
\bar B = {1 \over L} \int_\Gamma d \vec s \cdot \vec B (\vec x )
$$
and the curve $\Gamma$ is the line of sight (null geodesic)
of length $L$ from the quasar to the earth.

To compare (\ref{constraint}) with (\ref{faraday}), we need to know
something about the spectrum
of the magnetic field \cite{olesen}.
If the field is homogenous on the quasar distance
scale, then a direct comparison can be made and we see that the BBN
constraint is much weaker. This comparison is relevant if the magnetic
fields arise from inflation where it is assumed to be coherent on
horizon scales (subject to the discussion in point ({\em a}) above).
If, on the other hand, a causal mechanism is found to generate
a magnetic field, the field is likely to be coherent on scales that
are much smaller than the horizon, say on a scale $\xi$.
For example, in the case of magnetic fields generated in cosmological
phase transitions, we expect \cite{olesen}
$$
B_{rms} \simeq \sqrt{L \over \xi} {\bar B}
$$
and the Faraday rotation constraint in (\ref{faraday})
translates into
$$
B_{rms} < \sqrt{L \over\xi} 10^{-9} \ G
$$
The BBN constraint (\ref{constraint})
can therefore be much stronger than the Faraday rotation
constraint if $L/\xi$ is large.

Finally we would like to project the results of this paper from a
different viewpoint. This is that the current astrophysical constraints
on magnetic fields constrain the large scale magnetic field
strength. That is, they say something about the Fourier
modes with wavelengths comparable to quasar distances. BBN, on the
other hand, provides
a complementary tool for constraining the magnetic field energy
density which includes contributions from all wavelengths. And
for causal mechanisms that generate magnetic fields, BBN is likely
to probe the magnetic field spectrum at the smallest wavelength
while the Faraday rotation measure is mostly sensitive to wavelengths
of order the quasar distance scale.

\vskip 1.5\baselineskip
\nin{\Large \bf Acknowledgements}
\vskip .5\baselineskip

\nin The authors wish to thank Lawrence Krauss for
discussions during the early stages of this work. P.K. and T.V.
thank the Aspen Center for Physics for hospitality while this work
was in progress.

\newpage
\nin{\Large \bf Figure Captions}
\vskip \baselineskip

\nin
Figure 1: The equilibrium neutron fraction $X_n^{eq}$
as a function
of the photon temperature $\tg$ for several
choices of the magnetic field evolution parameter $\zeta$.

\vskip \baselineskip

\nin
Figure 2: For $\zeta=2$, the neutron fraction $X_n$
near
the nucleosynthesis temperature region, ${\cal O}$(70) keV, is shown
for separate pieces
of the full calculation. The changes from
the electron thermodynamic functions,
(\rhoe, \Pe, \drhoedt\ ),
the weak interaction rates $\lambda_{n\leftrightarrow p}$, and
the magnetic field energy density $\rho_{B}$, on the neutron
fraction are shown individually.  Also plotted is the neutron
fraction with all/none of the above corrections.

\vskip \baselineskip

\nin
Figure 3: The neutron fraction $X_n$
as a function
of the photon temperature $\tg$ for several
choices of the magnetic field evolution parameter $\zeta$.
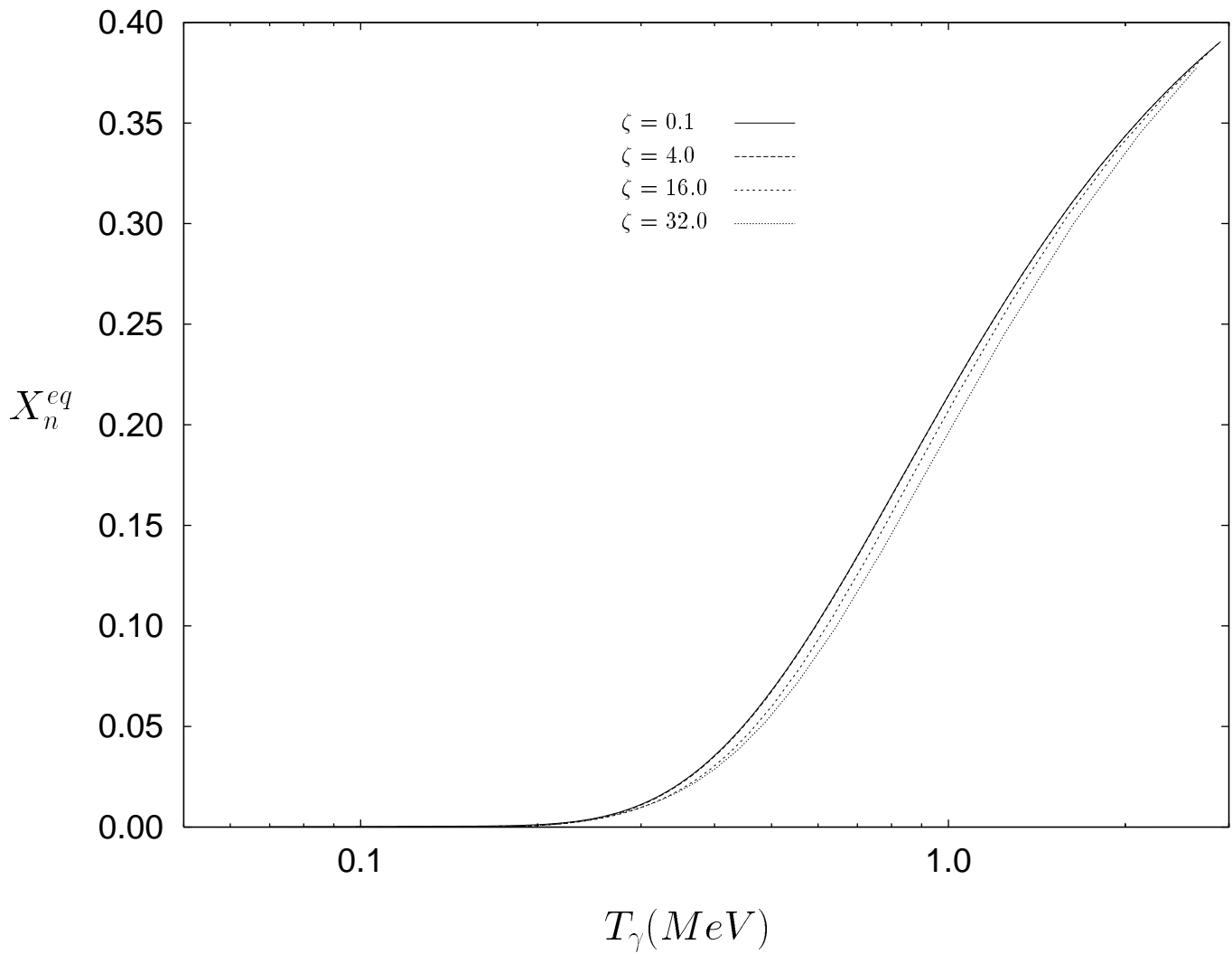
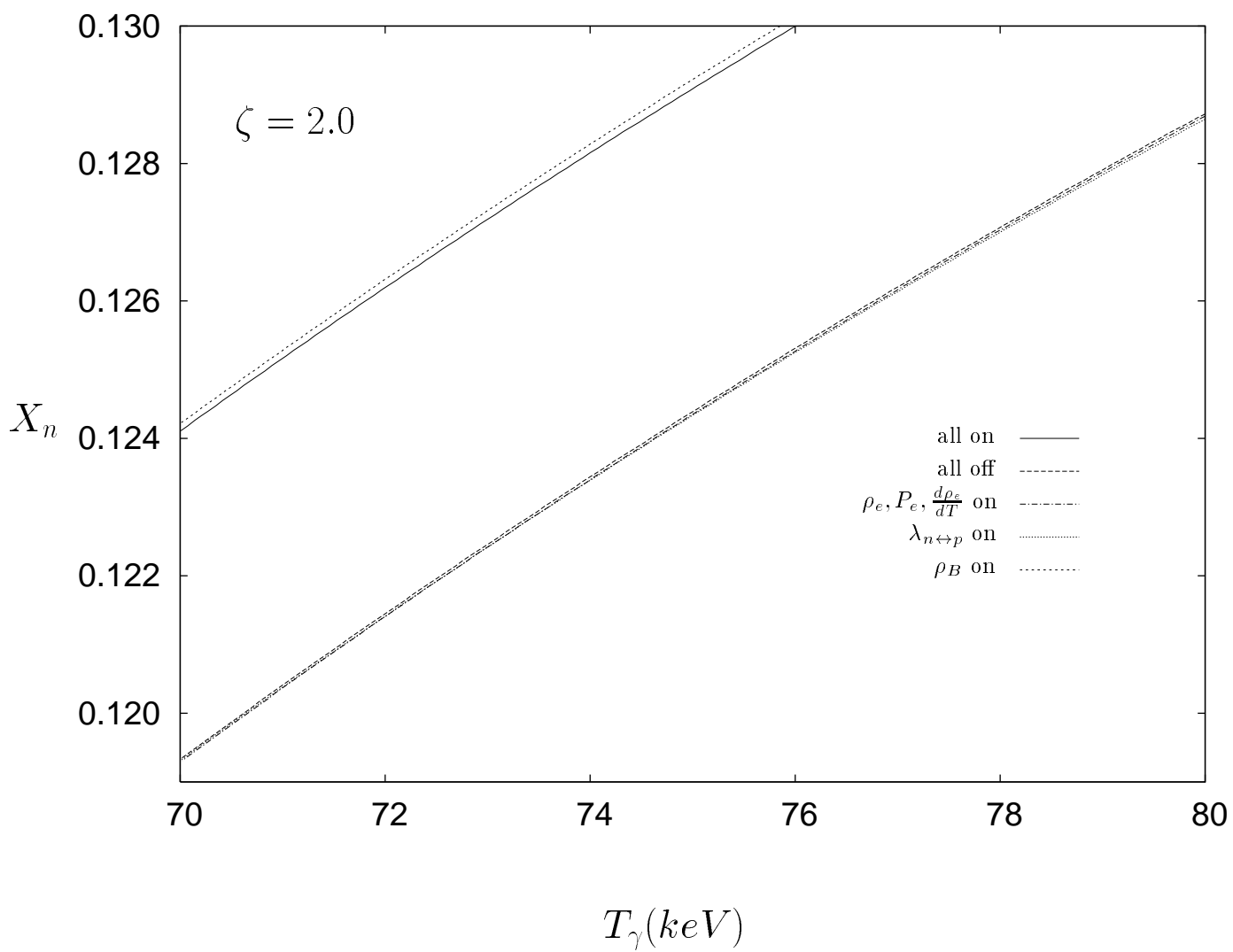
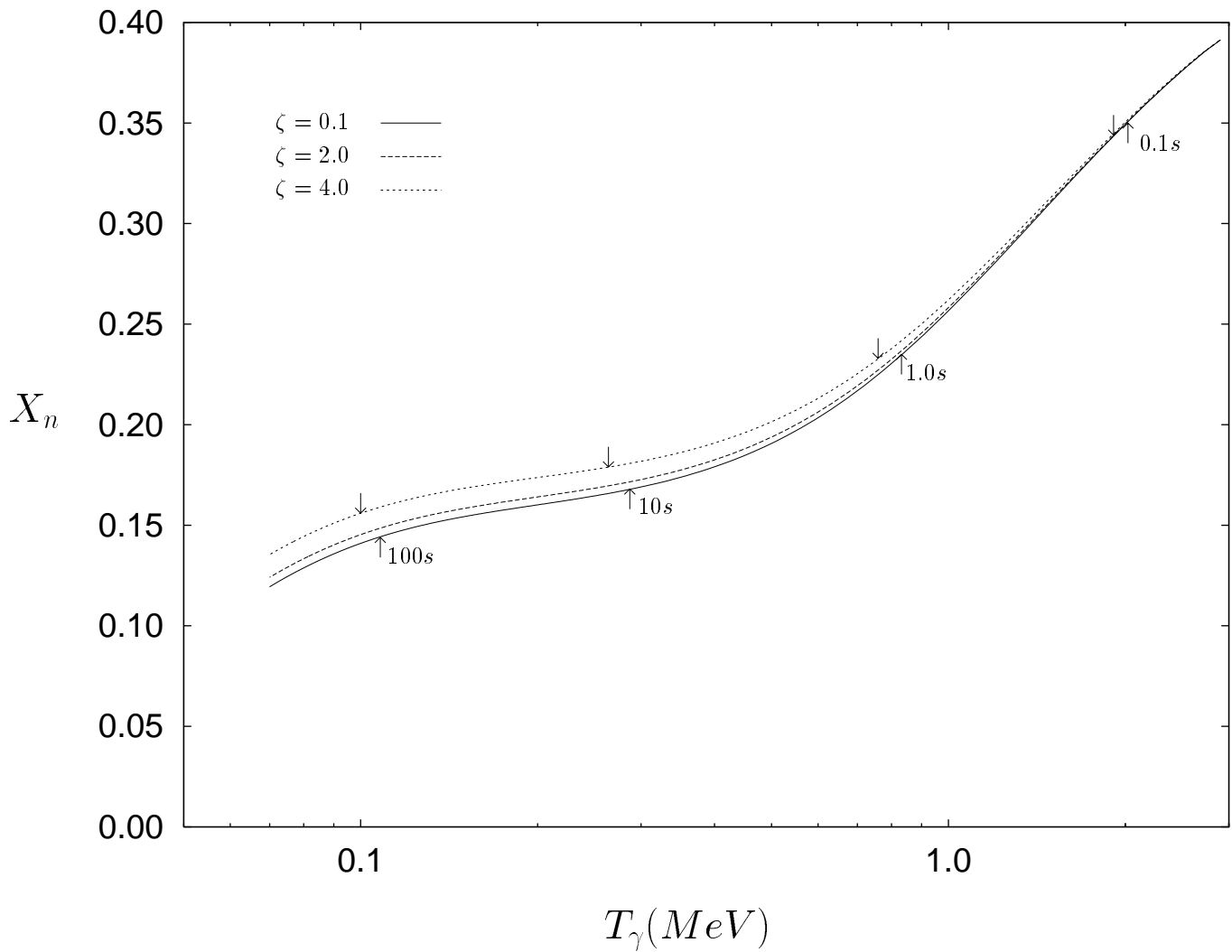
\end{document}